\documentclass[twocolumn]{aastex62}

\usepackage{amsmath}
\usepackage{amssymb}
\usepackage{placeins}
\usepackage{ulem}
\received{XXX}
\revised{XXX}
\accepted{XXX}
\submitjournal{ApJ}

\shorttitle{Dust-to-gas ratio resurgence in circumstellar disks.}
\shortauthors{Turrini et al.}

\begin{document}

\title{Dust--to--gas ratio resurgence in circumstellar disks due to the 
formation of giant planets: the case of HD\,163296}


\author{Turrini, D.}
\affiliation{INAF-Istituto di Astrofisica e Planetologia Spaziali (INAF-IAPS), Via del Fosso del Cavaliere n. 100, 00133, Rome, Italy}

\author{Marzari, F.}
\affiliation{Department of Physics and Astronomy, University of Padova, Via Marzolo 8, 35131 Padova, Italy}

\author{Polychroni, D.}
\affiliation{INAF-Istituto di Astrofisica e Planetologia Spaziali (INAF-IAPS), Via del Fosso del Cavaliere n. 100, 00133, Rome, Italy}

\author{Testi, L.}
\affiliation{INAF-Osservatorio Astrofisico di Arcetri, Largo E. Fermi 5, I-50125 Firenze, Italy}
\affiliation{European Southern Observatory (ESO), { Karl-Schwarzschild-Stra{\ss}e} 2, 85748 Garching bei { M\"{u}nchen}, Germany}
\affiliation{ Excellence Cluster Origins, Boltzmannstr. 2, D-85748 Garching, Germany}

\begin{abstract}
{
The amount of dust present in circumstellar disks is expected to steadily decrease with age due to the growth from $\mu m$--sized particles to planetesimals and planets. Mature circumstellar disks, however, can be observed to contain significant amounts of dust and possess high dust--to--gas ratios. Using HD\,163296 as our case study, we explore how the formation of giant planets in disks can create the conditions for collisionally rejuvenating the dust population, halting or reversing the expected trend. We combine N--body simulations with statistical methods and impact scaling laws to estimate the dynamical and collisional excitation of the planetesimals due to the formation of HD\,163296's giant planets. We show that this process creates a violent collisional environment across the disk that can inject collisionally produced second-generation dust into it, significantly contributing to the observed dust-to-gas ratio. The spatial distribution of the dust production can explain the observed local enrichments in HD\,163296's inner regions. The results obtained for HD\,163296 can be extended to any disk with embedded forming giant planets and may indicate a common evolutionary stage in the life of such circumstellar disks. Furthermore, the dynamical excitation of the planetesimals could result in the release of transient, non-equilibrium gas species like { H$_{2}$O, CO$_{2}$, NH$_{3}$ and CO} in the disk due to ice sublimation during impacts and, due to the excited planetesimals being supersonic with respect to the gas, could produce bow shocks in the latter that could heat it and cause a broadening of its emission lines.
} 
\end{abstract}



\section{Introduction} 
\label{sec:intro}

Many fundamental steps of the planetary formation process take place during the lifetime of circumstellar discs, among which are the settling of dust towards the median plane, the formation of planetesimals by dust accumulation, the growth of giant planets by planetesimal and gas accretion, and their possible orbital migration through interactions with the nebular gas (see e.g.\citealt{morbidelli2016} for a recent review). 

It is expected that both the density and size distribution of the original dust, present in the disk at the beginning of the settling process, will significantly change during the disk evolution time-scale \citep{testi2014}. The accretion of dust into planetesimals and planets will lead to a progressive decrease of its density with the disk's age \citep{pascucci2016}, in particular at small sizes. Thus, the dust depletion should peak at the time when the giant planets reach their final mass and finally clear the region from the remaining dust.

{ Across their formation, however, giant planets drastically alter the dynamical equilibrium of the surrounding 
planetesimals by exciting their orbits, 
 a process that acts in response to the mass growth of the giant planets independently of whether they migrate or not \citep{turrini2011,turrini2012,turrini2014a,turrini2014b,turrini2015,raymond2017,turrini2018}.} This phase of dynamical excitation {was shown to} greatly enhance the collisional activity among the planetesimals \citep{turrini2012}. 

The {resulting} energetic collisional evolution of the planetesimals, characterized by cratering and fragmentation events \citep{turrini2012}, {could in principle} reverse the process of dust depletion in circumstellar disks and {allow for} the dust--to--gas ratio to climb back up. The steady decline predicted on the basis of planetesimal and planet formation and drift towards the star would {therefore} see a halt followed by a sudden increase lasting as long as the planetesimal {impact rates remain sufficiently high}. 

{ A similar event, even if less dramatic, is invoked  {\it after} the circumstellar disks have lost their gaseous component and transitioned into debris disks to explain the increases in brightness of the latter. Differently from the case of circumstellar disks hosting growing giant planets, this \textit{delayed stirring} (see \citealt{wyatt2008} for a review) predicts that a planetesimal belt is stirred either by the secular perturbations of a nearby planet or due to a period of dynamical instability, generally assuming that the planetesimal population is in a collisional steady state (e.g. \citealt{weidenschilling2010,thebault2012,kral2016}).}

\subsection{HD\,163296 as a benchmark disk}

{ A potential and promising test bench to study the dust rejuvenation process while it is ongoing is the circumstellar disk surrounding HD\,163296.} { ALMA's Cycle 2 and 4 observations of HD\,163296's circumstellar disk, with a spatial resolution of 25 and 4 au respectively, showed distinct gaps in the dust distribution of the disk, suggesting the presence of at least three giant planets \citep{isella2016,isella2018,dullemond2018}}. These observations {suggest} they orbit approximately at 60, 105 and 160\,au from the central star {(based on HD\,163296's pre-Gaia distance from the Sun of 122 pc, see below for more details)} and allowed for constraining their fiducial masses to 0.1, 0.3 and 0.3 Jovian masses, albeit with large uncertainties \citep{isella2016}. 

{ Thanks to more refined numerical modelling with independent techniques, these mass values have been recently revised upward to 0.46, 0.46 and 0.58 Jovian masses \citep{liu2018}, with the masses of the two outer giant planets being proposed to be as large as { \mbox{ 1 Jovian mass}} \citep{teague2018}. In parallel, the presence of a fourth giant planet, with mass of about 2 Jovian masses and orbiting at about 260 au from the star, has also been proposed \citep{pinte2018}.} 

{ \citet{isella2016} detected the presence of dust from the innermost, not resolved regions of HD\,163296's disk up to 250 au, { with the gas extending twice as far from the star and reaching about 500 au, and reconstructed the surface density profiles of both dust and gas}. If one assumes an inner edge of the disk at 0.1 au, integrating the dust surface density profile reconstructed by \citet{isella2016} up to 250 au yields $\sim$420 M$_{\oplus}$ of dust grains. Conversely, integrating the gas surface density profile reconstructed by \citet{isella2016} up to 500 au and assuming a gas--to--dust ratio of 100:1 as in the interstellar medium \citep{bohlin1978,andre2000,lada2007,natta2007} yields an expected dust mass of $\sim$280 M$_{\oplus}$. HD\,163296's disk therefore appears to contain 1.5 times the amount of dust  expected for its current gaseous mass or, equivalently, to possess an overall gas--to--dust ratio of $\sim$67.}

{Before the release of the { second data release (DR2)} catalogue { \citep{gaia2018}} of the ESA space mission {{Gaia}} { \citep{gaia2016}}, HD\,163296 was characterized as an intermediate mass star of 2.3 M$_{\odot}$ with a distance from the Sun of 122 pc and an age of about  5\,Myr \citep{vandenancker1997}. Following {{Gaia}}'s observations, HD\,163296's distance has been revised downward to 101.5 pc {(\citealt{bailer-jones2018}, based on the astrometric data from {{Gaia}}'s DR2 catalogue in \citealt{gaia2018})}: this change results in a revised mass for the star of 1.9 M$_{\odot}$ and in a more compact system where all planetary orbits should be scaled accordingly. Nevertheless, both the pre- and post-{{Gaia}} values indicate that the system is evolved and characterized by the coexistence of dust, gas, planetesimals and planets.} 

{Since its features suggest that HD\,163296 should have already undergone or even still be undergoing the dynamical excitation phase caused by the mass growth of its giant planets, in this paper 
we explore the dynamical excitation of planetesimals 
for the different proposed values of its planetary masses, and 
test if their enhanced collisional evolution can lead to a significant production of second-generation dust 
and raise the dust--to--gas ratio of this system to the observed value.}

\section{Numerical Methods} 
\label{sec:methods}

{Our investigation is based on the combination of N--body simulations, aimed at assessing the dynamical excitation caused in HD\,163296's disk by the formation of the giant planets, with statistical methods to estimate the impact fluxes and impact velocities among the planetesimals and scaling laws for the outcomes of collisions in the different impact regimes, with the goal of providing a first assessment of the implication of the dynamically excited environment on the collisional production of dust.}

\subsection{Modelling the dynamical excitation process}\label{sec:dynamical_model}

{ The N--body simulations were performed using \textit{Mercury-Ar$\chi$es}, a parallel implementation of the hybrid symplectic algorithm of the MERCURY 6 software from \citet{chambers1999} that also allows for including gas drag, orbital migration and planetary mass growth in the simulations. 

The simulations considered a set of HD163296's analogues composed of the central star, the three forming giant planets initially reported by \citet{isella2016} and supported by the independent analyses of \citet{liu2018} and \citet{teague2018}, and a disk of planetesimals modelled with $10^{5}$ massless particles. In this study we did not include the presence of the fourth, outer planet suggested by \citet{pinte2018} due to its still poorly constrained orbital and physical characteristics. Nevertheless, we will briefly discuss its expected impact on our results when drawing the conclusions of this study.} 

{In order to ease the comparison with previous studies \citep{isella2016,liu2018,teague2018}, and particularly with HD\,163286's gas and dust distributions \citep{isella2016}, { following \citet{teague2018} we adopted HD\,163296's pre-{{Gaia}} distance and the planetary semimajor axes from pre-{{Gaia}}'s works \citep{isella2016,liu2018} in the simulations and in the discussion of their outcomes}.} 

{The planetesimal disk we considered in this study extended from 10 au (i.e. well inside the orbital region resolved by the observations of \citealt{isella2016}) to 250 au (i.e. the outer border of the dust distribution reconstructed by \citealt{isella2016}). The orbital regions corresponding to the feeding zones of the giant planets (e.g. \citealt{dangelo2011} and references therein) were left empty as planetesimals originally there would be incorporated into the growing giant planets. 

{ Similarly to \citet{turrini2012}, the initial orbits of the planetesimals were characterized by values of eccentricity and inclination (in radians) uniformly distributed between 0 and $10^{-2}$ \citep{weidenschilling2008}. As discussed in \citet{weidenschilling2011}, this choice of initial  conditions is equivalent to assuming a velocity dispersion between the planetesimals of the same order of the escape velocities from the largest planetesimals embedded in the swarm ($\sim$150 m/s, see Sect. \ref{sec:collisional_model} for details on the largest planetesimals considered). As we will show in Sect. \ref{sec:results}, the forced eccentricities and inclinations created by the dynamical excitation process are more than an order of magnitude higher, so our results are limitedly affected by their initial values.}

{At the radial distances from the star considered here (tens of au and larger) the gas drag is expected to have negligible effects on the dynamics of planetesimals \citep{weidenschilling1985}, particularly on a timescale of a few Myrs (i.e. the age of the HD\,163296 system) and for planetesimals with sizes of the order of ten km or larger. Nevertheless, we have included its effects for completeness.}

{The effects of the gas on the dynamics of the planetesimals were estimated by computing the drag acceleration $F_D$ (see \citealt{brasser2007} and references therein):
\begin{equation}
F_{D} = \frac{3}{8}\frac{C_{D}}{r_{p}}\frac{\rho_{g}}{\rho_{p}}v_{r}^{2}
\label{eqn-gasdrag}
\end{equation}
where $C_{D}$ is the gas drag coefficient, $\rho_{g}$ is the local density of the gas, $\rho_{p}$ and $r_{p}$ are the density and radius of the planetesimals respectively, and $v_{r}$ is the relative velocity of the planetesimals and the gas. The gas drag coefficient $C_D$ of each planetesimal is computed following the treatment described by \citet{brasser2007} as a function of the Reynolds number, the Mach number and the Knudsen number. This means that the individual gas drag coefficients are coupled both to the specific orbit of each planetesimal and to the local disk environments crossed during said orbit.}

{The local disk environments crossed by the planetesimals are characterized using the gas density and temperature profiles of HD\,163296's disk as reconstructed by \citet{isella2016}. In particular, the gas density profile adopted in the simulations is:
\begin{equation}
\Sigma(r) = \Sigma_{0}\left(\frac{r}{165\,au}\right)^{-0.8} \exp\left[-\left(\frac{r}{165\,au}\right)^{1.2}\right]
\label{eqn-gasdensity}
\end{equation}
where $\Sigma(r)$ is the radial profile of the total gas surface density and $\Sigma_{0} = 5.42$ g$\,$cm$^{-2}$. 

{ Following \citet{bergin2013}, Supplementary Information, the latter value is computed as $\Sigma_{0} = 2.37 \cdot \Sigma_{0}(^{12}CO)/(14\cdot n(^{12}CO/H_{2}))$ where $\Sigma_{0}(^{12}CO) = 1.6\times10^{-3}$ g$\,$cm$^{-2}$ is the measured value of the $\Sigma_{0}$ parameter for the $^{12}$CO surface density \citep{isella2016}, $n(^{12}CO/H_{2})=5\times10^{-5}$ is the $^{12}CO:H_{2}$ cosmic molecular abundance \citep{isella2016}, 14 is the ratio of the molecular weights between $^{12}CO$ and $H_{2}$, and $2.37$ is the mean molecular weight of the gas including, alongside hydrogen, also helium and all heavy elements \citep{bergin2013}.}

{For the planetesimals, in our reference simulations we adopted values of $r_{p} = 50$ km, the characteristic size of planetesimals formed by pebble accretion (e.g. \citealt{klahr2016}), and $\rho_{p} = 1$ g cm$^{-3}$, as a compromise between the measured densities of comets (0.4-0.6 g cm$^{-3}$, see e.g. \citealt{brasser2007} and references therein and \citealt{jorda2016}) and that of the larger ($\approx200$ km in diameter) ice-rich captured trans-neptunian object Phoebe (1.63 g cm$^{-3}$, \citealt{porco2005}). We also performed test simulations with $r_{p} = 5$ km and {\mbox{$r_{p} = 0.5$ km}} but we found negligible changes for all planetesimals whose orbits remain outside of 20-30 au (i.e. in the observationally resolved region of the { disk's density profiles from \citealt{isella2016}).}  

{In our simulations we focused on the \emph{in situ} formation scenario, in which the giant planets do not undergo any significant migration during their formation, to avoid including too many free parameters in the study.} The initial orbits of the giant planets were therefore characterized by semimajor axes identical to those estimated by \citet{isella2016} {and \citet{liu2018}} for the centres of the gaps, {and were assumed to be coplanar on the disk midplane} and with initial eccentricities of the order of $10^{-3}$ { to account for the damping effects of the tidal gas drag on the growing planetary cores \citep{cresswell2008}}. 

{The formation of the giant planets was assumed to occur on a relatively short timescale \citep{lambrechts2012,bitsch2015} and their mass growth was modelled using the numerical approach from \citet{turrini2011}.} During the first $\tau_{c}=10^{6}$ years of the simulations the giant planets accreted their cores, whose masses grew from an initial value of $M_{0}=0.1\,M_{\oplus}$ to the critical value $M_{c}=15\,M_{\oplus}$ as:
\begin{equation}
 M_{P}=M_{0}+\left( \frac{e}{e-1}\right)\left(M_{c}-M_{0}\right)\left( 1-e^{-t/\tau_{c}} \right)
\label{eqn-coregrowth}
\end{equation}
consistently with the mass growth profiles in previous studies of Jupiter's formation (see e.g. \citealt{lissauer2009} and \citealt{dangelo2011} and references therein) and in the pebble accretion scenario \citep{bitsch2015}.

After the critical mass value $M_{c}$ was reached, the mass growth of each giant planet during the subsequent gas accretion phase was modelled as:
\begin{equation}
 M_{P}=M_{c}+\left( M_{F} - M_{c}\right)\left( 1-e^{-(t-\tau_{c})/\tau_{g}}\right)
\label{eqn-gasgrowth}
\end{equation}
where $M_{F}$ is its final mass. An e-folding time of $\tau_{g}=1\times10^5$ years was chosen based on the results of the hydrodynamical simulations described in \citet{lissauer2009} and in \citet{coradini2010} and \citet{dangelo2011} and references therein.

{We performed three different simulations to estimate how the end results are affected by the current uncertainties on the masses of the giant planets{, as summarized in Table \ref{table-planets}}. In the first simulation, representing our reference case, the final masses for the giant planets were identical to the ones estimated by \citet{liu2018} (see { Table \ref{table-planets} and} Sect. \ref{sec:intro}). In the second simulation{, representing our ``low mass'' case,} the giant planets grew to the fiducial masses estimated by \citet{isella2016} (see { Table \ref{table-planets} and} Sect. \ref{sec:intro}). Finally, in the third simulation{ representing our ``high mass'' case,} we adopted the mass estimated by \citet{liu2018} for the innermost planet and 1 Jovian mass for the outer two following the results of \citet{teague2018}} { (see Table \ref{table-planets} and Sect. \ref{sec:intro})}.

\begin{table}
\centering
\begin{tabular}{cccc}
\hline
{ Scenario} & \multicolumn{3}{c}{{ Planetary Masses (in Jovian masses)}} \\
\hline
                  & \textit{Inner Planet}  & \textit{Central Planet}   & \textit{Outer Planet} \\
                  & ($a$ = 60 au) & ($a$ = 105 au) & ($a$ = 160 au) \\
\hline
``Low mass''  & 0.10 & 0.30 & 0.30 \\
Reference     & 0.46 & 0.46 & 0.58 \\
``High mass'' & 0.46 & 1.0  & 1.0  \\
\hline
\end{tabular}
\caption{Summary of the final planetary masses (in Jovian masses) adopted for the three giant planets in the three scenarios explored in this work. 
For reference, for each giant planet we also indicated the orbital semimajor axis (\textit{a}) adopted in the simulations.}\label{table-planets}
\end{table}

The orbital elements of the giant planets and the massless particles were recorded every $10^{6}$ years. The output of the simulations was used to study the evolution of the circumstellar collisional environment in response to the mass growth of the giant planets. For this task we took advantage of the well-tested collisional methods that have been extensively used to study the evolution of the asteroid belt in the Solar System (see e.g. \citealt{obrien2011} and references therein, \citealt{turrini2012}) and { have been applied} also to the study of debris disks \citep{weidenschilling2010}.

\subsection{Modelling the collisional dust production}\label{sec:collisional_model}

{Our collisional model is based on the numerical algorithm originally developed by \citet{wetherill1967} and expanded by later works (see \citealt{greenberg1988,farinella1992} and \citealt{obrien2011} and references therein)} to calculate the evolution {of the \emph{intrinsic impact probabilities} $P_{i}$ and of the distribution of the} \emph{impact velocities} $v_{i}$ among the planetesimals across the circumstellar disk due to their dynamical excitation. 

{From the individual intrinsic impact probabilities $P_{i}$ so estimated we computed the average intrinsic impact probabilities ${P}_{av}$ for each 1 au-wide ring between 10 and 250 au. 
From these average intrinsic impact probabilities it is then possible to compute the number of impacts occurring within each 1 au-wide ring during a given timespan $\Delta\,t$ using the following equation (see e.g. \citealt{obrien2011}):
\begin{equation}
N_{i} = {P_{av}} \, \left(R_{t}+R_{i}\right)^{2} \, N_{t} \, N_{i} \, \Delta t
\label{eqn-impacts}
\end{equation}
where $P_{av}$ is the average intrinsic impact probabilities ${P}_{av}$ of the specific 1 au-wide ring { considered} (measured in impacts per km$^{2}$ yr$^{-1}$, see \citealt{obrien2011} and references therein), $R_{t}$ and $R_{i}$ are the radii of the target body and the impactor respectively (the term between parentheses in Eq. \ref{eqn-impacts} being the total collisional cross-section of target and impactor, the term $\pi$ being incorporated into $P_{av}$ as discussed in \citealt{obrien2011} and references therein), $N_{t}$ and $N_{i}$ are the numbers of target bodies (within the 1 au-wide ring) and impactors (in the whole disk) with those specific sizes, and $\Delta\,t=10^{6}$ years based on the  outputs of the N--body simulations.}

{The number of target bodies and potential impactors for the different combinations of $R_{t}$ and $R_{i}$ can be estimated, for a given size-frequency distribution of the planetesimals, using the following equation \citep{weidenschilling2010}:
\begin{equation}
N = \int N(m) = C\,m^{-\gamma}\,dm
\label{eqn-N}
\end{equation}
where $\int N(m)$ is the number of planetesimals with masses in a given mass range, $\gamma$ is the exponential slope of the size-frequency distribution, and the constant $C$ links the total mass $M_{tot}$ to the mass contained in the specified mass range (whose lower and upper boundaries are $m_{min}$ and $m_{max}$) as \citep{weidenschilling2010}:
\begin{equation}
C = \frac{\left(2-\gamma\right)M_{tot}}{m^{2-\gamma}_{max}-m^{2-\gamma}_{min}}
\label{eqn-C}
\end{equation}
}

To compute the values of $N$ and $C$ using Eq. \ref{eqn-N} and \ref{eqn-C}, we need to constrain the unknown initial total mass and size-frequency distribution of the planetesimals embedded in the disk. 

{ To estimate the initial total mass of the planetesimal disk we adopted the following approach. We assumed that the original circumstellar disk of HD\,163296 had an initial mass equal to $20\%$ the mass of the star, i.e. that it was a few times more massive than it is now. This assumption is consistent both with the measured decay time of the gas in disks (2.3--3 Myr, see \citealt{fedele2010} and \citealt{ercolano2017}) and the observed high mass loss rates of HD\,163296 due to molecular wind (\citealt{klaassen2013}, see also \citealt{ercolano2017}).  We also assumed that its { initial} overall gas--to--dust ratio{, inherited from the molecular cloud, }was $100:1$ as measured in the interstellar medium \citep{bohlin1978,andre2000,lada2007,natta2007,ercolano2017}.} 

{If this mass was efficiently converted into planetesimals (e.g. by pebble accretion, consistently with the fact that HD\,163296 was capable of forming three giant planets at such distances from the star{; see however Sect. \ref{sec:results} for a discussion of the implications of a less efficient conversion}), the total mass of the original planetesimal disk amounts to about 1530 M$_\oplus$. From this value we subtracted the mass needed to form the three cores of the giant planets (i.e. 45 M$_\oplus$): { this leaves $M_{tot}=1475$\,M$_\oplus$ of planetesimals, which we adopted as our starting value. 

To put this value in the right context, it is important to point out two things. First, the measured abundance of dust in HD\,163296 amounts to 420 M$_\oplus$ (see Sect. \ref{sec:intro} and \citealt{isella2016}): since dust represents only the \emph{visible fraction} of the solid mass embedded in the disk, this mass value is a lower limit for the total solid mass in HD\,163296. Our adopted initial mass of the planetesimal disk is equivalent to assuming that this currently visible mass of dust represents about $30\%$ of the total amount of solid material in the protoplanetary disk. Second, our results on the dust production scale linearly with the mass of the planetesimal disk (see Eq. \ref{eqn-impacts} and the dependance on the number of targets in each ring), so that an initially less massive planetesimal disk will simply reduce the amount of produced dust proportionally.
}

\begin{figure}[t]
\centering
\includegraphics[width=\columnwidth]{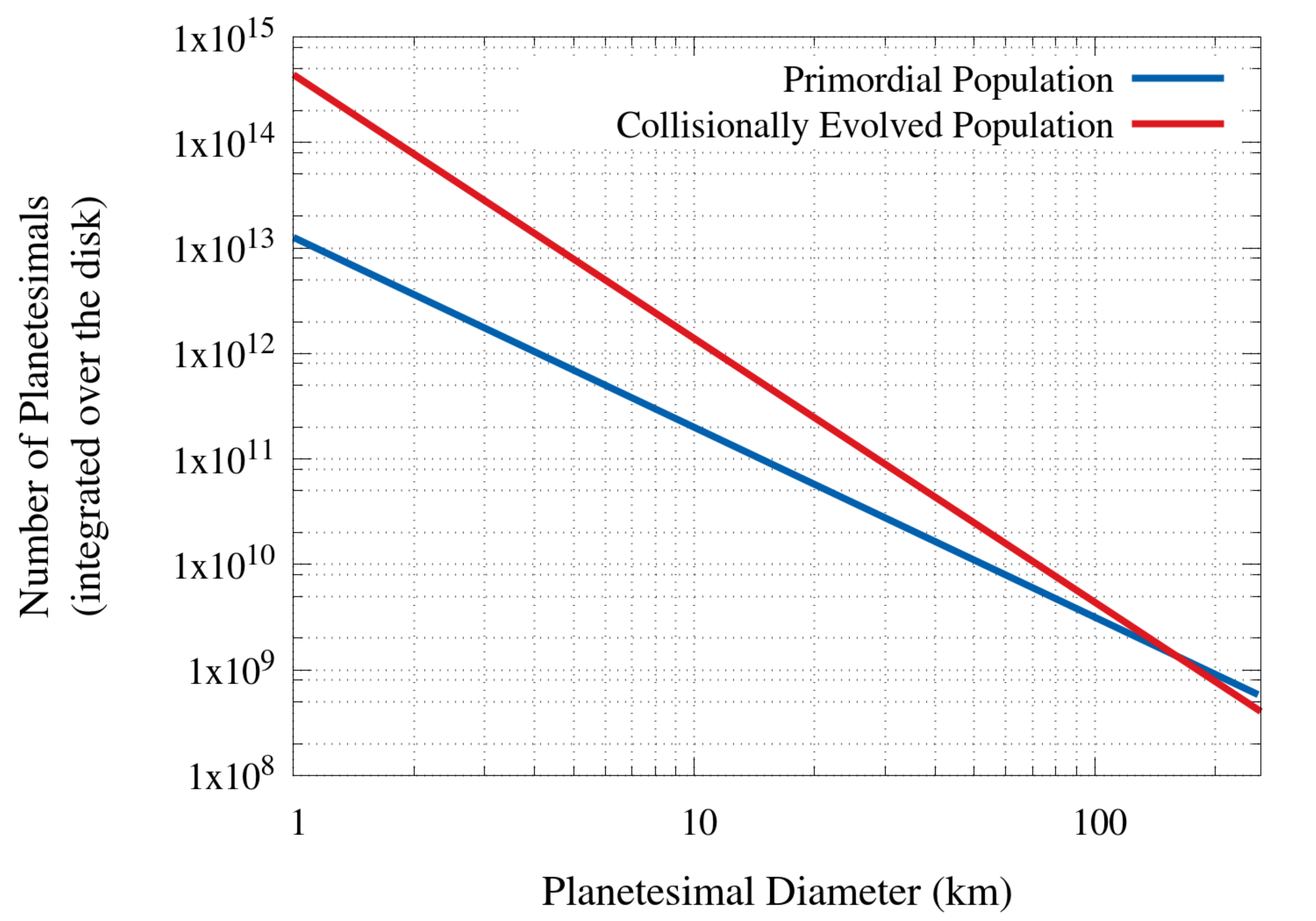}
\caption{Comparison of the disk-integrated populations of planetesimals predicted by the primordial and the collisionally evolved size-frequency distributions in the range of planetesimal diameters where they are both defined (see main text for details).}
\label{figure-sfd}
\end{figure}

{For what it concerns the exponential slope of the size-frequency distribution of the planetesimals, we considered two different cases that are based on the study of debris disks by \citet{krivov2018}. The first one is the \emph{primordial size-frequency distribution} expected for a young population of planetesimals formed by pebble accretion, characterized by an exponent $\gamma=1.6$ in Eqs. \ref{eqn-N} and \ref{eqn-C} (see \citealt{krivov2018} and references therein). The second one is a \emph{collisionally evolved size-frequency distribution} expected for a population of planetesimals in a collisional steady-state (see e.g. \citealt{weidenschilling2010,krivov2018}), characterized by an exponent $\gamma=11/6$ in Eqs. \ref{eqn-N} and \ref{eqn-C}. A comparison of the two size-frequency distributions is shown in Fig. \ref{figure-sfd}.}

{For both size-frequency distributions we followed \citet{krivov2018} and adopted the upper cut--off size of $d_{max} = 400$\,km that, for the assumed planetesimal density $\rho_{p} = 1$\,g cm$^{-3}$, is equivalent to adopting $m_{max} = \left(\pi / 6 \right) \, \rho_{p} \, d_{max}^{3} = 3.35 \times 10^{22}$\,g. Because of the different nature of the two size-frequency distributions, we adopted separate lower cut-off sizes for each of them. For the primordial size-frequency distribution, we adopted the lower cut--off size of $d_{min} = 1$\,km (see \citealt{krivov2018} and references therein) equivalent to $m_{min} = 5.24 \times 10^{14}$\,g. For the collisionally evolved size-frequency distribution, we adopted the lower cut--off size of $d_{min} = 1$\,m equivalent to $m_{min} = 5.24 \times 10^{5}$\,g. For both size-frequency distributions we then proceeded to bin the planetesimals so that each bin would contain planetesimals with diameters comprised between $d_{i}$ and $\sqrt{2}\,d_{i}$ (e.g. between 1 and 1.4\,km or between 16 and 22.6\,km, \citealt{catwg1979}).
}

{To estimate the effects of the impacts over the expected wide range of impact conditions (both in terms of impact velocities and sizes of the involved bodies), instead of the piece-wise collisional model adopted in \citet{turrini2012} we took advantage of the scaling law recently derived by \citet{genda2017} and valid both in the regime of cratering erosion and catastrophic disruption:
\begin{equation}
\frac{m_{ej}}{m_{tot}}=0.44 \phi \times max(0,1-\phi) + 0.5 \phi^{0.3} \times min(1,\phi)
\label{eqn-erosion}
\end{equation}
where $m_{ej}$ is the fraction of mass ejected during the impact averaged over all possible impact angles, $m_{tot}$ is the sum of the impactor mass $m_{i}$ and the target mass $m_{t}$, and $\phi$ is the ratio among the specific impact energy $Q$ and the critical specific impact energy $Q_{D}^{*}$. Following \citet{genda2017}, we adopted $m_{ej}/m_{tot} = 1$ when $\phi \geq 10$.}

{We defined $Q$ as:
\begin{equation}
Q = \frac{1}{2} \mu v_{imp}^{2}/m_{tot}
\end{equation}
where $v_{imp}$ is the impact velocity and $\mu$ is the reduced mass of the impactor-targer pair $\left(m_{i} m_{t}\right)/\left(m_{tot}\right)$ \citep{genda2017}. Following \citet{krivov2018}, we adopted different prescriptions for defining $Q_{D}^{*}$ for the two size-frequency distributions due to the different expected interior state of the planetesimals.}

{For the primordial size-frequency distribution, characterized by loosely bound planetesimals mainly held together by self-gravity in the size range $d_{min} - d_{max}$ considered (see \citealt{krivov2018} and references therein), for each planetesimal with diameter $d_{i}$ we computed $Q_{D}^{*}$ (in erg/g, see \citealt{krivov2018}) as:
\begin{equation}
Q_{D}^{*} = 7\times10^{4} \left( \frac{0.5 \, d_{i}}{r_{0}} \right)^{-1.59} \left( \frac{v_{i}}{v_{0}} \right)^{0.5} + \frac{6 \, G \, m}{5 \, d_{i}}
\end{equation}
where $r_{0} = 1$\,mm and $v_{0}=3$\,km s$^{-1}$ \citep{krivov2018}.
}

{For the collisionally evolved size-frequency distribution, characterized by monolithic planetesimals (see e.g. \citealt{weidenschilling2010,krivov2018} and references therein), for each planetesimal with diameter $d_{i}$ we computed $Q_{D}^{*}$ (in erg/g, see \citealt{krivov2018}) as:
\begin{align}
Q_{D}^{*} = &  \, 5\times10^{6} \left( \frac{0.5 \, d_{i}}{r_{0}} \right)^{-0.36} \left( \frac{v_{i}}{v_{0}} \right)^{0.5} \nonumber \\
& + 5\times10^{6} \left( \frac{0.5 \, d_{i}}{r_{1}} \right)^{1.38} \left( \frac{v_{i}}{v_{0}} \right)^{0.5}
\end{align}
where $r_{0} = 1$\,m, $r_{1} = 1$\,km and $v_{0}=3$\,km s$^{-1}$ \citep{krivov2018}.
}

{Each planetesimal of size $d_{i}$ was considered as a potential target for all planetesimals with size equal or smaller, and as a potential impactor for all planetesimals with size equal or greater. Given the exploratory nature of this study, in our collisional model we did not track directly the production of dust but made the simplifying assumption that $20\%$ of the ejected mass resulting from Eq. \ref{eqn-erosion} is in the form of dust, here loosely defined { as grains up to the order of cm in size, with the bulk of the mass contained in the larger grains (from $\sim$0.1 mm to cm in size, \citealt{okeefe1985})}. From a physical point of view, this can be interpreted as assuming that all smaller fragments produced by impacts (e.g. those near or below our lower cut-offs in size) get efficiently converted into dust within the time resolution of our collisional model (i.e. $10^{6}$ years), which is broadly consistent with the results of more complex collisional models (see e.g. \citealt{weidenschilling2010,krivov2018}).}

{We also did not track dynamically the changes in the population of planetesimals of different sizes as a result of collisions. These changes (e.g. the enrichment of the population of smaller planetesimals due to the ejection of collisional fragments or its depletion due to the growth by larger planetesimals) should result in the gradual transition from our primordial size-frequency distribution to our collisionally evolved one (see also \citealt{weidenschilling2010,krivov2018} and references therein). From a realistic point of view, therefore, at any given time after the formation of HD\,163296's giant planets the expected dust production should be located somewhere between the one associated to the primordial size-frequency distribution and the one associated to the collisionally evolved size-frequency distribution.}

\begin{figure*}
\centering
\includegraphics[width=\textwidth]{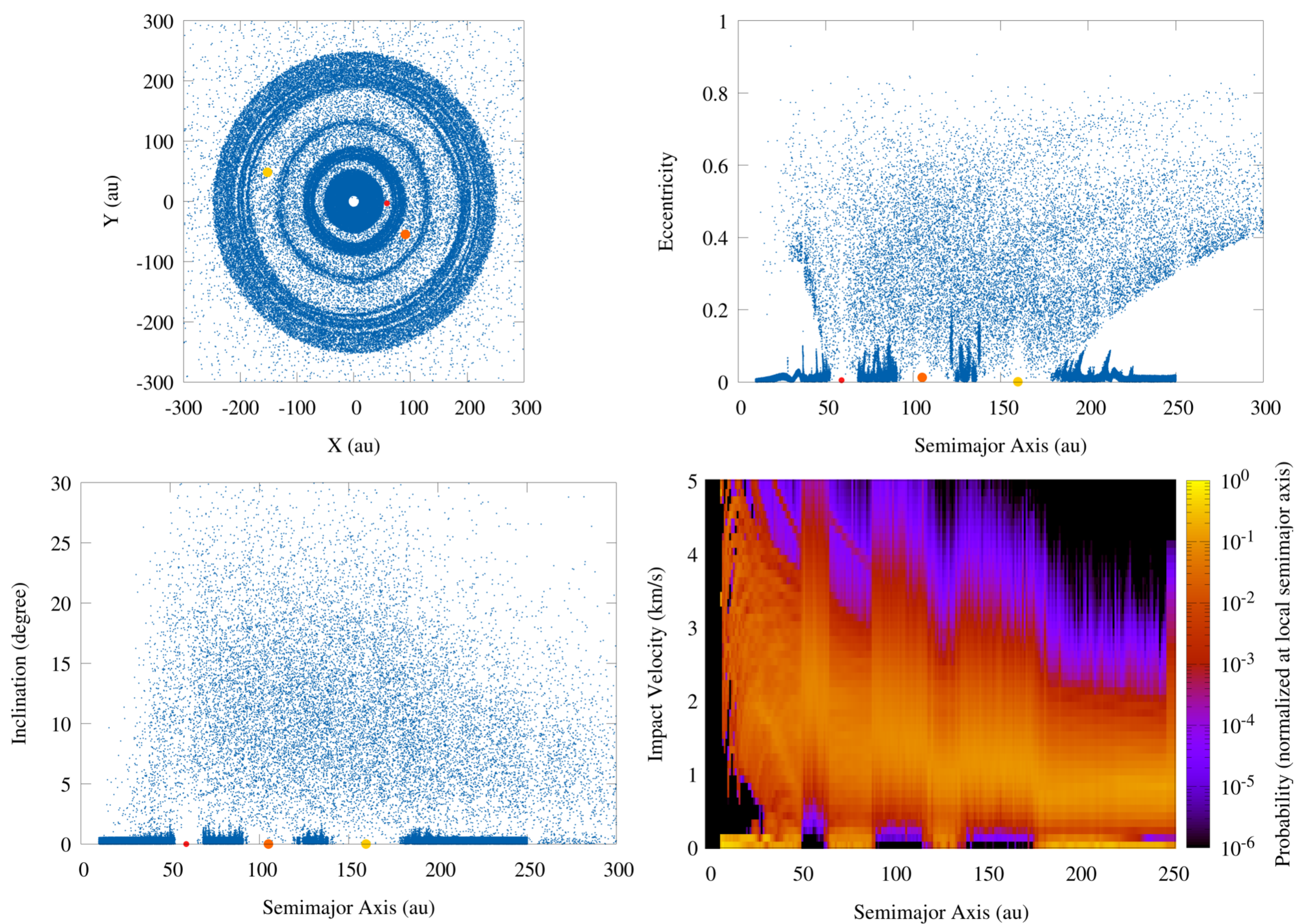}
\caption{Dynamical state of the planetesimal disk of HD\,163296 in our reference case (i.e planetary masses as estimated by \citealt{liu2018}) after 5 My due to the excitation caused by its three giant planets. Top left: ``face-on'' orbital structure of the planetesimal disk. Top right: orbital eccentricities of the planetesimals in the excited circumstellar disk.  Bottom left: orbital inclinations of the planetesimals in the excited circumstellar disk. Bottom right: radial distribution of the impact velocities among planetesimals throughout the excited circumstellar disk. The color code indicates the probability distribution of the impact velocities normalized at the local semimajor axis. This means that that each vertical slice of the plot represents the impact velocity distribution for planetesimals at that specific semimajor axis.}\label{figure-summary}
\end{figure*}

\section{Results}
\label{sec:results}

{In presenting the results, we will  first describe the excited dynamical environment created by the formation of the giant planets and its implications for the collisional environment. In doing so, we will first describe the general picture depicted by our reference scenario (i.e. the one where the planetary masses are those estimated by \citealt{liu2018}) and then discuss the differences with the ``low mass'' and ``high mass'' scenarios. Finally, we will present the results of our simplified collisional model for the dust production in HD\,163296's system.}

\subsection{The dynamical excitation process in HD\,163296's disk: reference scenario}

{Fig. \ref{figure-summary} summarizes the state of the simulated system in our reference scenario after 5\,Myr of dynamical evolution, i.e. a possible present state for HD\,163296's planetesimal disk. As is immediately visible, the gravitational perturbations of the giant planets carved not only the observed gaps in the gas and/or the dust \citep{isella2016,isella2018} but also analogous gaps in the planetesimal disk (see Fig. \ref{figure-summary}, top panels and bottom left panel) creating a population of scattered planetesimals on highly eccentric and/or inclined orbits (Fig. \ref{figure-summary}, top right and bottom left panels).} 

In parallel, the appearance of giant planets created a network of orbital resonances across the disk through which they dynamically excited the orbits of the planetesimals outside the gaps (see Fig. \ref{figure-summary}, top right and bottom left panels). Both populations of dynamically excited bodies cross larger orbital regions than their non-excited counterparts and can impact against the latter at higher relative velocities than those characteristic of the initially unperturbed disk (see Fig. \ref{figure-summary}, bottom right panel, and Fig. \ref{figure-velocity}). 

{The temporal evolution of the dynamical excitation process in the semimajor axis vs. eccentricity plane is shown in Fig. \ref{figure-ae-time} while the corresponding temporal evolution of the impact velocities is shown in Fig. \ref{figure-velocity}. Both figures show snapshots of the dynamical state of the system { at 0 (i.e. the initial conditions of the simulations), 1, 2 and 5 Myr.} As can be immediately seen, the first 1 Myr (i.e. the time encompassing the first two top panels in Fig. \ref{figure-ae-time}) has limited effects on both the dynamical excitation of the planetesimals and their impact velocities. The only noteworthy change is the light increase in the impact velocities in the planetesimal ring between the two innermost planets due to the planetesimals excited by the growing planetary cores: impact velocities, however, remain sub-km/s.}
\begin{figure*}
\centering
\includegraphics[width=\textwidth]{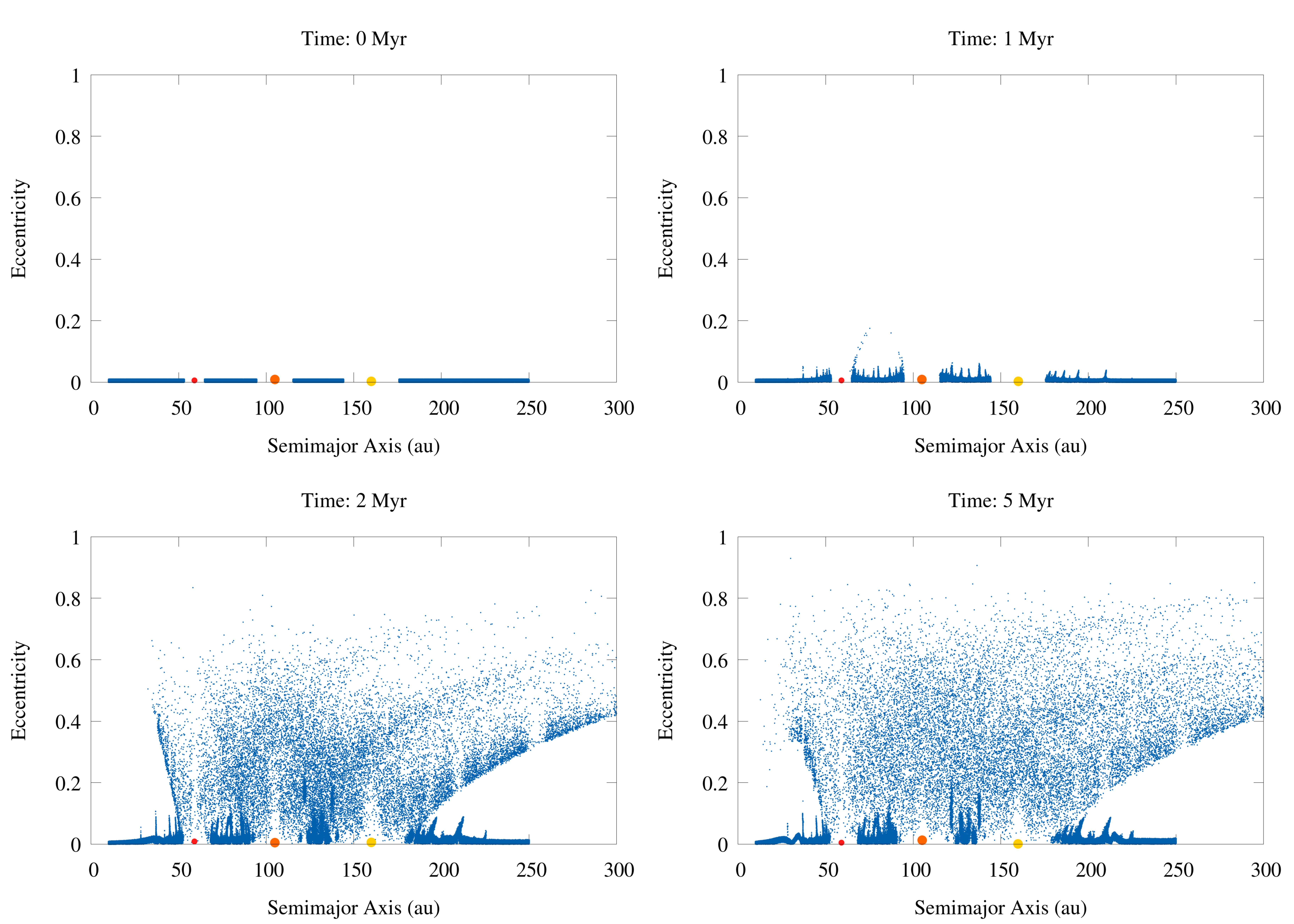}
\caption{Temporal evolution, in the semimajor axis--eccentricity plane, of the dynamical state of the planetesimal disk of HD\,163296 in our reference scenario (i.e planetary masses as estimated by \citealt{liu2018}). Going from left to right, top to bottom, the panels show the evolution from the initial state of the planetesimal disk in our simulations (top left panel) to its potential current state (bottom right panel).}\label{figure-ae-time}
\end{figure*} 
%
\begin{figure*}
\centering
\includegraphics[width=\textwidth]{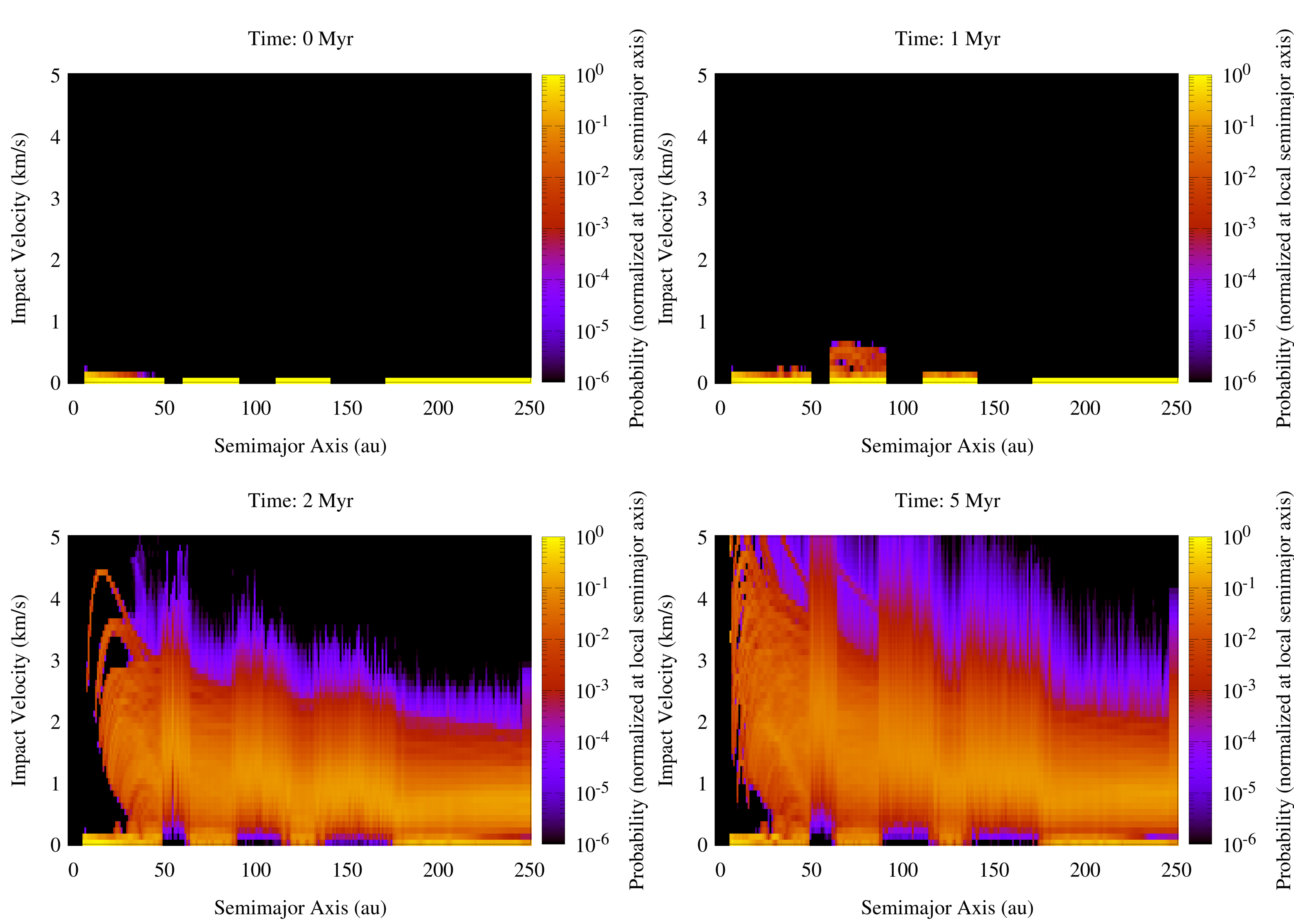}
\caption{Temporal evolution of the impact velocities across the planetesimal disk of HD\,163296 in our reference scenario (i.e planetary masses as estimated by \citealt{liu2018}). Going from left to right, top to bottom, the panels show the evolution of the impact velocity distribution from the initial state of the planetesimal disk in our simulations (top left panel) to its potential current state (bottom right panel). The color code indicates the probability distribution of the impact velocities normalized at the local semimajor axis. This means that that each vertical slice of the plot represents the impact velocity distribution for planetesimals at that specific semimajor axis.}\label{figure-velocity}
\end{figure*}
\begin{figure*}
\centering
\includegraphics[width=\textwidth]{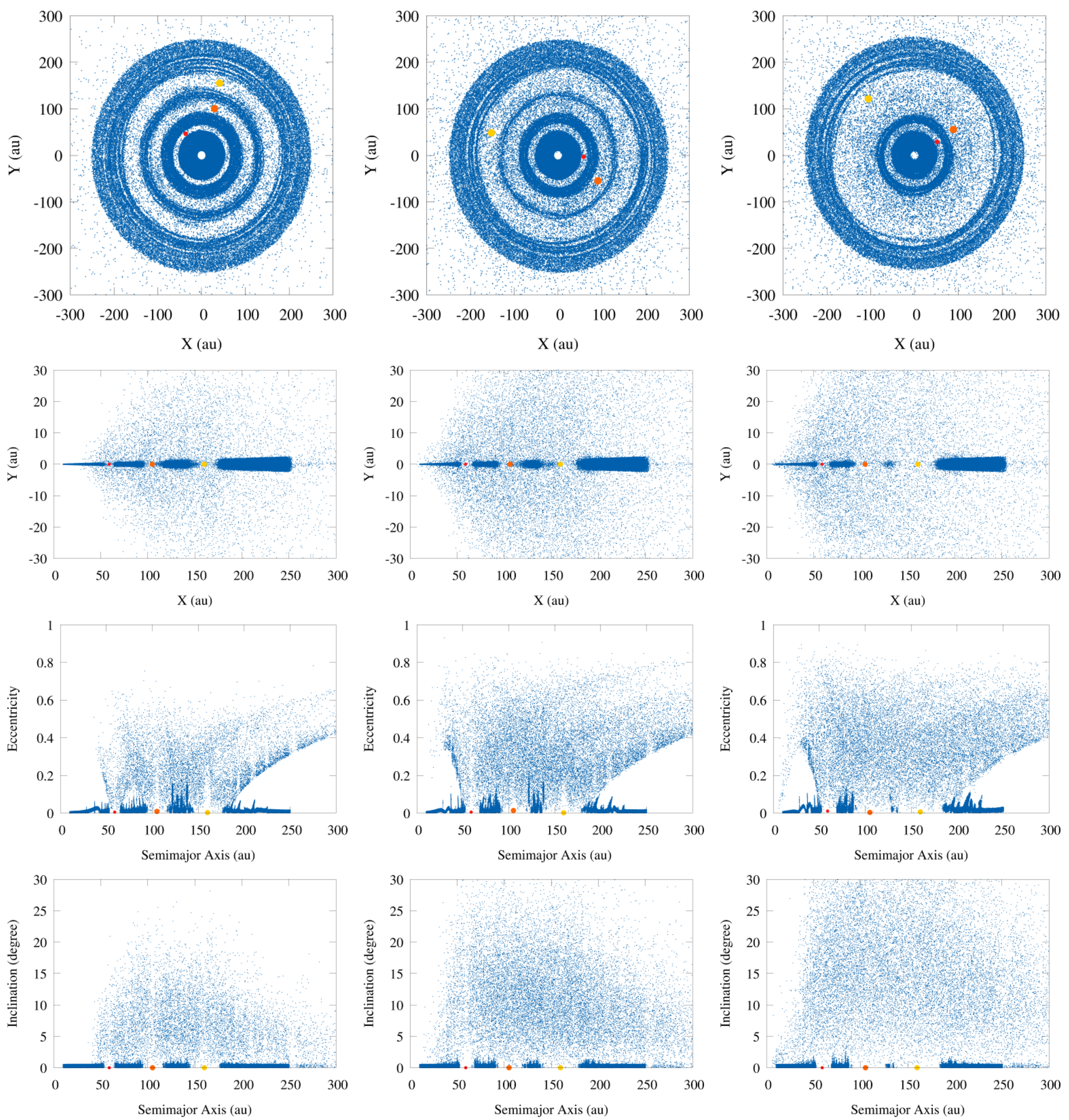}
\caption{Comparison of the orbital structure and dynamical excitation of HD\,163296's planetesimal population at 5 Myr in the three scenarios we considered for the mass values of the giant planets. From left to right, the plots show the ``low mass'' scenario (planets possessing the mass values estimated by \citealt{isella2016}), the reference scenario (planets possessing the mass values estimated by \citealt{liu2018}), and the ``high mass'' scenario (the two outer planets possessing the mass values estimated by \citealt{teague2018}). From top to bottom, the plots show the disk of planetesimals seen ``face on'' and ``edge on'', the disk of planetesimals in the semimajor axis--eccentricity plane, and the disk of planetesimals in the semimajor axis--inclination plane.}\label{figure-comparison}
\end{figure*}


{During the second Myr (i.e. the time between the top right and the { bottom} left panels in Figs. \ref{figure-ae-time} and \ref{figure-velocity}) the giant planets rapidly grow to their final masses by gas accretion, significantly affecting both the dynamical state of the planetesimal disk and the distribution of the impact velocities. The planetesimal disk suddenly acquires an extensive population of dynamically excited bodies (see Fig. \ref{figure-ae-time}) and the impact velocities grow up to 2-4\,km/s throughout its whole radial extension (see Fig. \ref{figure-velocity}). { During the following 3\,Myr} the dynamical excitation not only does not start decreasing, but is actually slowly continuing to build up (see Figs. \ref{figure-ae-time} and \ref{figure-velocity}, { bottom right panels}), with the highest impact velocities reaching and exceeding 5\,km/s.}  

\subsection{Mass-dependence of the dynamical excitation: comparing the different scenarios}

\begin{figure}
\centering
\includegraphics[width=\columnwidth]{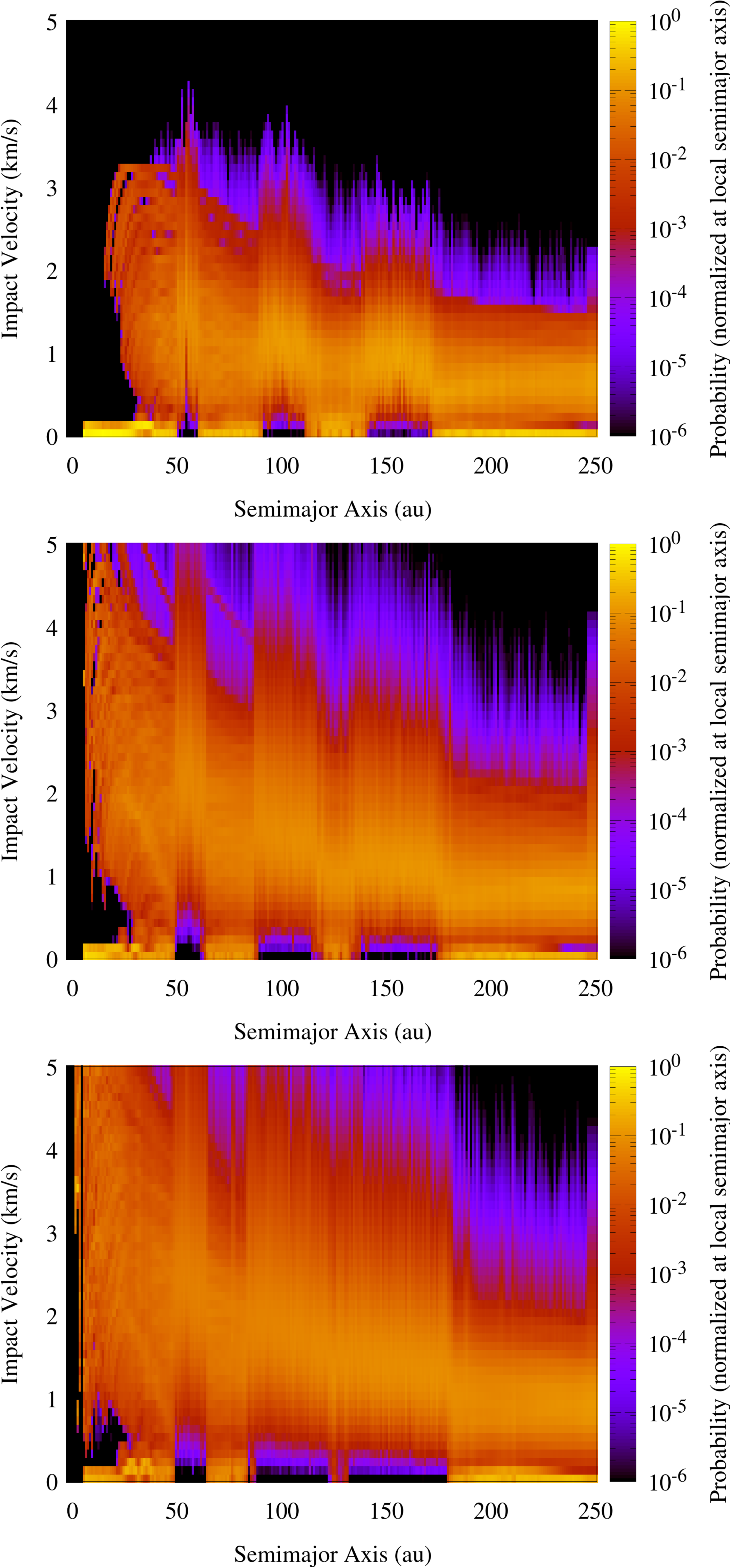}
\caption{Comparison of the impact velocities across the planetesimal disk of HD\,163296 at 5 Myr in the three scenarios we considered for the mass values of the giant planets. From top to bottom the plots show the ``low mass'' scenario (planets possessing the mass values estimated by \citealt{isella2016}), the reference scenario (planets possessing the mass values estimated by \citealt{liu2018}), and the ``high mass'' scenario (the two outer planets possessing the mass values estimated by \citealt{teague2018}). The color code indicates the probability distribution of the impact velocities normalized at the local semimajor axis. This means that that each vertical slice of the plot represents the impact velocity distribution for planetesimals at that specific semimajor axis.}\label{figure-velocity-comparison}
\end{figure}

{

As discussed in Sect. \ref{sec:intro}, the estimated masses of the giant planets embedded in HD\,163296's disk are { still uncertain, which is why we set out to explore the dynamical excitation process in different scenarios for the planetary masses as summarized in Table \ref{table-planets}}. Figs. \ref{figure-comparison} and \ref{figure-velocity-comparison} show the differences in the orbital structure and dynamical excitation, and in the distribution of the impact velocities of HD\,163296's population of planetesimals in the three scenarios we considered. 

As can be immediately seen, from a qualitative point of view the picture previously described when discussing the reference scenario holds also in the ``low mass'' and ``high mass'' scenarios. The giant planets always create a population of dynamically excited planetesimals with high eccentricities and/or high inclinations. The highest impact velocities are always recorded inside the innermost planet, while the lowest impact velocites are always in the outer part of the planetesimal disk, beyond the orbit of the outermost planet.

From a quantitative point of view, however, there is a number of significant differences among the three scenarios. While the number of surviving massless particles, i.e. the dynamical tracers of the planetesimals in the N--body simulations, vary limitedly ($99\%$ in the ``low mass'' scenario, $96\%$ in the reference scenario, $86\%$ in the ``high mass'' scenario), their spatial distribution and dynamical characteristics change significantly. 

Specifically, in Fig. \ref{figure-comparison} one can see that the well--defined rings of planetesimals visible in the ``low mass'' scenario get thinner for increasing planetary masses, with the ring comprised between the two outermost planetesimals disappearing in the ``high mass'' scenario. Since only a fraction of the original planetesimals is dynamically ejected from the system even in the ``high mass'' scenario, this means that the planetesimals originally orbiting inside the rings became part of the dynamically excited population of planetesimals on high--eccentricity and/or high--inclination orbits. 

This is showcased by the bottom half part of Fig. \ref{figure-comparison}, where the orbital elements of the planetesimals are shown in the semimajor axis vs. eccentricity and semimajor axis vs. inclination planes. The maximum orbital eccentricity values grow from about 0.6 in the ``low mass'' scenario to about 0.8 in the reference and ``high mass'' scenarios. Similarly, the distribution of the bulk of the orbital inclination values grows from 0--20$^{\circ}$ to 0--30$^{\circ}$.

These changes in the dynamical excitation and orbital characteristics of the planetesimals have a direct impact on the distribution of the impact velocities. Fig. \ref{figure-velocity-comparison} shows the comparison between the three scenarios for the planetary masses. As for the dynamical excitation, also for the impact velocities one can immediately see a linear growth with increasing planetary masses. 

Specifically, in the ``low mass'' scenario the bulk of the excited impact velocities clusters between 0.5--1\,km/s and the highest impact velocities fall between 3 and 4\,km/s (see Fig. \ref{figure-velocity-comparison}). In the reference and the ``high mass'' scenario, instead, the bulk of the excited impact velocitites clusters between 1--2\,km/s while the highest impact velocities reach and exceed 5\,km/s (see Fig. \ref{figure-velocity-comparison}). 

}

\subsection{Collisional dust production by HD\,163296's planetesimal population}

\begin{figure}
\centering
\includegraphics[width=\columnwidth]{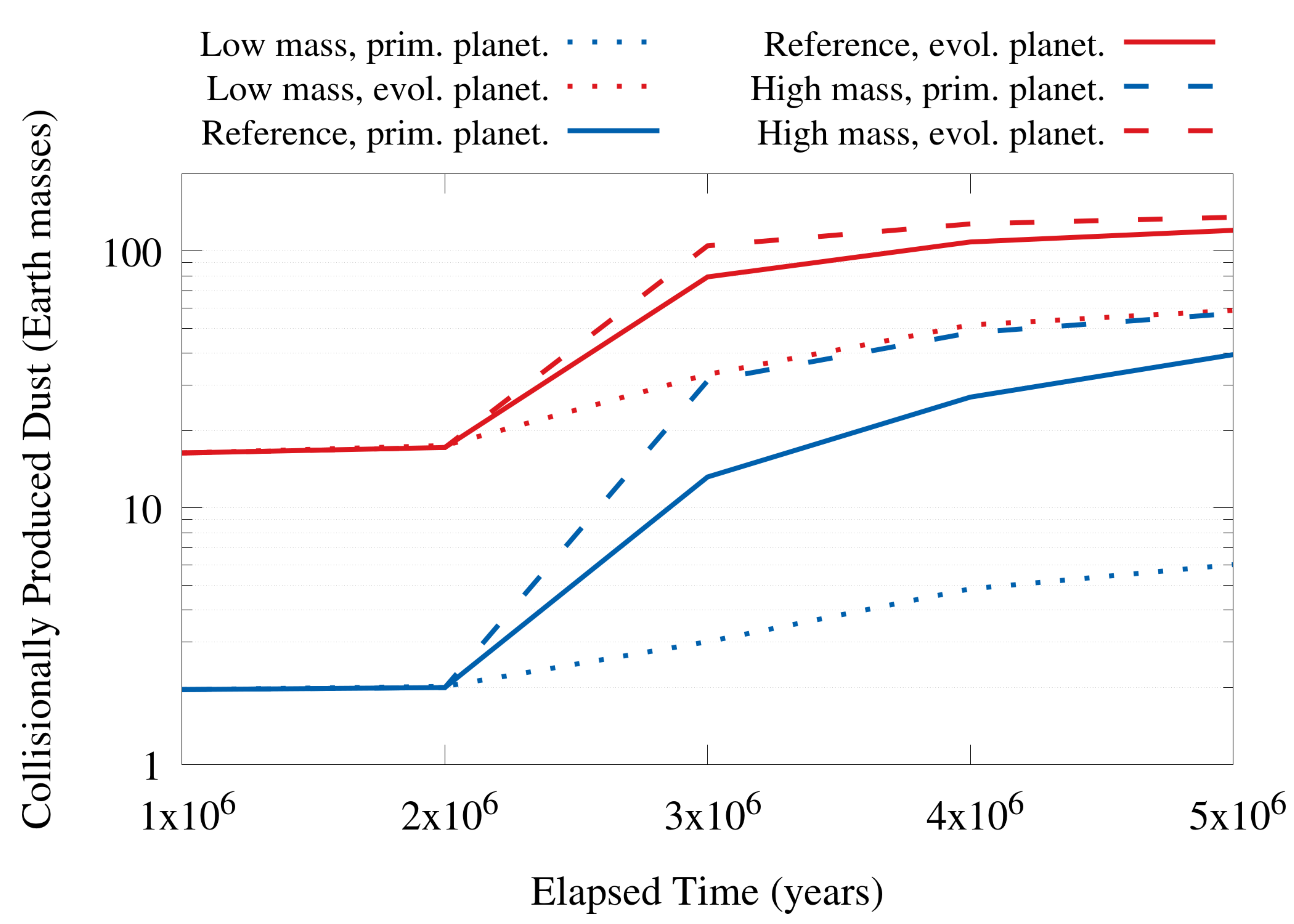}
\caption{Comparison between the global dust production (in $M_{\oplus}$) of the planetesimal collisional evolution over each 1 Myr of life of HD\'163296's system in the three scenarios for the planetary masses (``low mass'' scenario: dotted lines; reference scenario: solid line; ``high mass'' scenario: dashed line) and for the two size-frequency distributions of the planetesimals we considered (primordial size-frequency distribution: blue line; collisionally evolved size-frequency distribution: red line).}\label{figure-dust-time}
\end{figure}
\begin{figure}
\centering
\includegraphics[width=\columnwidth]{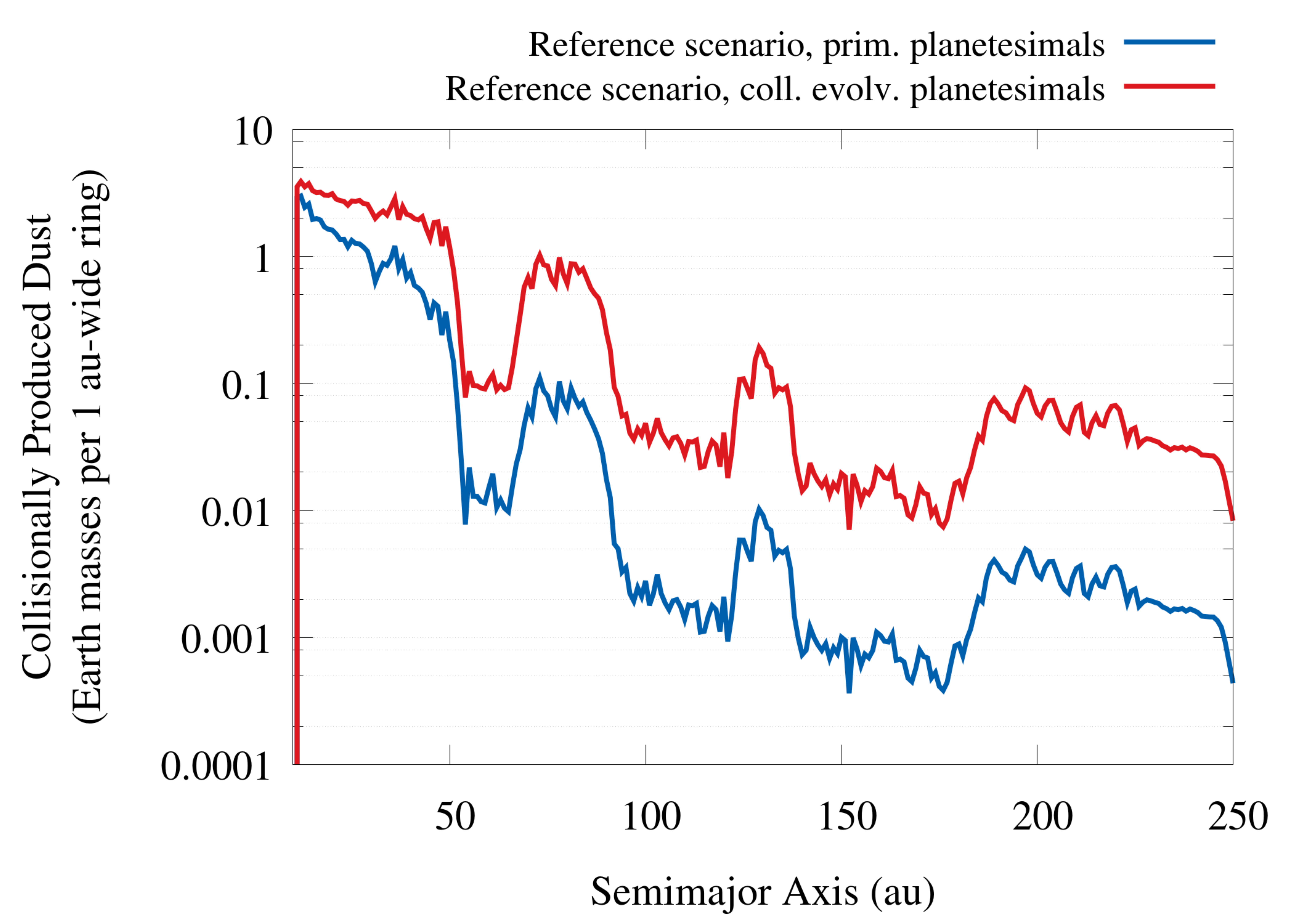}
\caption{Example of the local dust production (in $M_{\oplus}$ per 1 au-wide ring) between 4 and 5 Myr in our reference scenario (i.e. assuming the planetary masses estimated by \citealt{liu2018}) and for both size-frequency distributions we considered for the planetesimals (primordial size-frequency distribution: blue line; collisionally evolved size-frequency distribution: red line).}\label{figure-dust-space}
\end{figure}

\begin{table}
\centering
\begin{tabular}{ccc}
\hline
{Scenario} & \multicolumn{2}{c}{{Dust Production (in $M_{\oplus}$)}} \\
\hline
                  & \textit{Primordial}     & \textit{Collisionally} \\
                  & \textit{SFD}            & \textit{Evolved SFD}   \\
\hline
``Low mass''  & 24  & 228 \\
Reference     & 123 & 448 \\
``High mass'' & 200 & 523 \\
\hline
\end{tabular}
\caption{Cumulative dust production (in $M_{\oplus}$) due to the planetesimal collisional evolution over the life of HD\,163296's disk in the three scenarios for the planetary masses and for the two size-frequency distributions (SFDs in the table) of the planetesimals we considered. { For reference, by integrating the dust surface density profile reconstructed by \citet{isella2016} the total observed dust amounts to about 420 M$_\oplus$.}}\label{table-dust}
\end{table}

{

The total dust production associated to the planetesimal collisional evolution, integrated over the whole life of HD\,163296's disk, is reported in Table \ref{table-dust}. As can be immediately seen, the results are significantly different between the two size-frequency distributions adopted for the planetesimals, with the dust production varying by almost a factor of 10 in the case of the primordial size-frequency distribution while in the case of the collisionally evolved size-frequency distribution the difference between the minimum and maximum dust production is only slighly more than a factor of two.

This difference arises from the different dominant sources of dust in the two cases. In the case of the collisionally evolved size-frequency distribution, the main source of dust is provided by the efficient cratering erosion of the abundant small planetesimals, particularly at sub-km sizes where the impact rates become significantly higher (see e.g. \citealt{turrini2012}). On the contrary, the primordial size-frequency distribution contains most mass in the form of large ($\geq$100 km) planetesimals and is comparatively deficient of smaller planetesimals (see Fig. \ref{figure-sfd}). 

The primordial size-frequency distribution becomes capable of producing significant amounts of dust only when the impact velocities are high enough to allow for the break-up of such large objects, cratering erosion being far less efficient at these sizes due to the higher escape velocities \citep{turrini2012}. Due to their weaker internal structure, however, when the required impact velocities are reached the impacts between these large planetesimals can inject large quantities of dust into the disk, explaining the larger growth observed in Table \ref{table-dust}.

As shown in Table \ref{table-dust}, when the planetesimal population is characterized by a collisionally evolved, steady state size-frequency distribution, its dynamical excitation and collisional evolution in response to the formation of the giant planets appear capable of injecting enough dust to explain $50\%$ of the observed dust { (i.e. 420 M$_\oplus$, see Table \ref{table-dust} and Sect. \ref{sec:intro})} even in the ``low mass'' scenario. In the reference and in the ``high mass'' scenarios, the collisional production of dust appears capable of explaining all of the currently observed dust.

As discussed above, the case of a planetesimal population characterized by a primordial size-frequency distribution produces different results. In the ``low mass'' scenario, the amount of produced dust would be limited (about 20 $M_{\oplus}$) and would not affect the global dust abundance in any significant way. In the reference and ``high mass'' scenarios, however, the collisional production of dust would be enough to contribute the extra $\sim$140 $M_{\oplus}$  of dust seen when comparing the integrated gas and dust profiles discussed in Sect. \ref{sec:intro}.

It should be noted that one of the assumptions of our simplified collisional model was that all the mass initially present as dust in the circumstellar disk was efficiently converted into planetesimals before the giant planets reached their present masses. To explain the observed dust abundance in the case of a primordial size-frequency distribution of the planetesimals, however, between half and two-thirds of the currently observed dust should be primordial, accordingly reducing the original mass of the planetesimal disk. 

If the primordial dust present in HD\,163296's disk amounts to $280$ M$_{\oplus}$ (as estimated from integrating the current gas density profile { and scaling it by its initial dust--to--gas ratio}, see Sect. \ref{sec:intro}), the numbers reported in Table \ref{table-dust} for the primordial size-frequency distribution of the planetesimals should be reduced by about $20\%$, meaning that the collisional production of dust would amount to $100-160$ M$_{\oplus}$ in the reference and ``high mass' scenarios. Therefore, collisional dust production could still explain the global overabundance of dust discussed in Sect. \ref{sec:intro} if the planetary masses are at least equal to those estimated by \citet{liu2018}.

Fig. \ref{figure-dust-time} shows the temporal evolution of the collisional production of dust over time. The first two Myrs contribute only marginally to the injection of second--generation dust into the circumstellar disk, which is instead produced in the three Myrs following the formation of the giant planets. The planetesimal formation process can therefore continue undisturbed until the giant planets reach their final masses. This means, in turn, that the assumption on the efficient conversion of the primordial dust into planetesimals before the onset of the dynamical excitation process at the basis of our simplified collisional model is physically realistic.

Finally, in Fig. \ref{figure-dust-space} we show the collisionally produced dust between 4 and 5\,Myrs in the reference scenario as an example of the radial profile of the injection of second--generation dust in the disk. While a detailed description of the spatial distribution and production of the second---generation dust is beyond the scope of our simplified collisional model, it is worth noting that the peak of dust production naturally occurs within the orbit of the innermost planet as a combination of the higher spatial density of the planetesimals and of the higher impact velocities (see Figs. \ref{figure-velocity} and \ref{figure-velocity-comparison}). 

This is particularly interesting since the orbital region within the innermost planet is the one where \citet{isella2016} found the largest discrepancy in the dust--to--gas ratio between the observations and the theoretical expectations based on the results of hydrodynamic simulations of the evolution of gas and dust. Specifically, the innermost planet was expected to act as an effective barrier to the inward diffusion of dust from the outer regions of the disk { and to cause this orbital region to become more and more depleted of dust over time. On the contrary, this region shows some of the highest dust--to--gas ratios in the whole disk \citep{isella2016}.}

Integrating the dust profile from \citet{isella2016} reveals a dust content of about 90 M$_{\oplus}$ between 20 and 50 au. { However, based on the hydrodynamic simulations performed by \citet{isella2016} and the low average dust--to--gas ratio (1:200) they predict in this orbital region, one would expect a dust content of only about 20 M$_{\oplus}$. Our simplified collisional model reveals that the collisional dust production can supply the missing 70 M$_{\oplus}$ of dust} in 1-2\, Myrs in the case of the collisionally evolved size-frequency distribution (in all scenarios for the planetary masses) and in 2-3\, Myrs in the case of the primordial size-frequency distribution (in the reference and ``high mass'' scenarios).

{ We performed a similar comparison between the recently estimated masses of the dust rings located between each consecutive pair of giant planets (i.e. 50-60 M$_\oplus$ for the ring between the innermost and central planets and 40-45 M$_\oplus$ for the ring between the central and the outermost planets, \citealt{dullemond2018}) and the dust production in the same orbital regions. 

Our comparison reveals that the collisional dust production can supply all the mass contained in the innermost dust ring and about 10-20$\%$ the mass contained in the outermost dust ring in 2-3 Myr in the case of the collisionally evolved size-frequency distribution. In the case of the primordial size-frequency distribution the collisional dust production can explain only up to 10$\%$ of the observed mass of the rings. Our results would therefore seem to indicate that the dust population in the outer regions of HD\,163296's disk is characterized by a mixture of primordial and second-generation dust.

The inclusion of the fourth giant planet proposed by \citet{pinte2018}, however, should increase the dynamical excitation in the outer regions of the planetesimal disk and locally enhance both the impact velocities and the dust production. This, combined with the radial drift and trapping of the dust \citep{dullemond2018}, could result in a better match with the estimated ring masses, particularly for the primordial size-frequency distribution due to its higher sensitivity to the impact velocity.

}

\subsection{Additional environmental effects of the dynamical excitation and collisional evolution}

Due to the range of values spanned by the enhanced impact velocities, the amount of material stripped from the planetesimals by impacts is not the only factor affected by the process of dynamical excitation. Impact experiments on ice \citep{stewart2008} { as well as the observations of the Deep Impact mission to comet Tempel 1 \citep{ahearn2005}} reveal that also the physical state of the eroded material is affected. 

Impact velocities below 1\,km\,s$^{-1}$ are expected to cause the icy component of the planetesimals to be preferentially excavated instead of vaporized \citep{stewart2008}. As such, a large number of collisions (see Figs. \ref{figure-velocity} and \ref{figure-velocity-comparison}) in the dynamically excited disk will produce second-generation refractory and icy grains that will enrich the surviving first-generation original dust population of the disk. For impact velocities above 1\,km\,s$^{-1}$, impacts will melt and vaporize increasingly larger fractions of the icy component of the planetesimals \citep{stewart2008}. Through this process, the most energetic impacts (see Figs. \ref{figure-velocity} and \ref{figure-velocity-comparison}) will release in the disk gaseous species not in local thermal equilibrium with the surrounding gas, most notably { H$_{2}$O and CO$_{2}$ \citep{ahearn2005}, NH$_{3}$ and CO}.

While non-equilibrium species are expected to be transient and to freeze-out on relatively short timescales, it has been argued that collisions among planetesimals might sustain their continued presence beyond their respective ice condensation lines provided that impact rates are sufficiently high \citep{salinas2016}. Such scenario would be consistent with the possible detection of excess H$_{2}$O in the (unresolved) \emph{Herschel} observations of the circumstellar disk of HD\,163296 \citep{fedele2012} and would provide an explanation to the possible presence of both H$_{2}$O and NH$_{3}$ beyond their respective icelines in the circumstellar disk of TW Hya (also based on unresolved \emph{Herschel} observations, \citealt{salinas2016}).

{ Recent observations of DCO+ in the disk of HD\,163296 \citep{salinas2018} further support the possibility of an ongoing collisional release of non-equilibrium species beyond their respective snowlines. \citet{salinas2018} measured the presence of DCO+ beyond the CO show line, located at about 90 au, and their DCO+ measurements indicate a mostly constant abundance of this molecule between $\sim$90 and $\sim$180 au, followed by a prompt decline between $\sim$180 and $\sim$300 au. The author suggest this behaviour to be the result of the thermal desorption and/or photodesorption of moderate amounts of CO from the ice to the gas phase and its reaction with H$_2$D+ (the so-called cold deuteration channel, \citealt{salinas2018}) and discuss different mechanisms that can produce such localized desorption. 

While a detailed comparison between the DCO+ measurements and the collisional environment created by the giant planets is going to require dedicated studies, it is interesting to note that the region of higher DCO+ abundance beyond the CO snowline (i.e. $\sim$90-180 au, \citealt{salinas2018}) matches the orbital region excited be the second and third giant planets where the peak impact velocities reach 5 km/s (see Fig. \ref{figure-velocity-comparison}). The sharp decrease between 180 and 300 au occurs instead in the orbital region where the peak impact velocities do not surpass 3-3.5 km/s (see Fig. \ref{figure-velocity-comparison}). 

When the lower spatial density of planetesimals in the outer regions of the disk is taken into consideration, impact-driven desorption of CO appears qualitatively capable of  providing a viable explanation to the observed DCO+ trend. It is interesting to note that the fourth giant planet proposed by \citet{pinte2018} to be located at 260 au would increase the impact velocities in these outer orbital regions but would also lower the spatial density of the planetesimals (by increasing their orbital eccentricity and/or inclination), so that its presence does not necessarily invalidate the picture discussed above.}

A further environmental effect of the population of high eccentricity--high inclination planetesimals moving supersonically with respect to the gas is the {\it generation of shock waves} in the gas of the disk \citep{weidenschilling1998}. The high temperatures of the gas at the shocks may lead to the broadening of emission lines, which could be an observable test for the presence of supersonic planetesimals. In addition, according to \citet{tanaka2013} the heating and resulting evaporation of the planetesimal surfaces at bow shocks would also contribute to the release of gas species not in local thermal equilibrium with the surrounding gas. Finally, the subsequent cooling of the vapor produced in this way would form dust particles by recondensation, which would contribute to the formation of second-generation dust.

{
 Again, while a detailed treatment of the heating effects of supersonic planetesimals on the gas will require a future dedicated study, it is interesting to compare the morphology of the CO thermal emission reconstructed by \citet{isella2018} with the orbital structure of the planetesimal disk created by the dynamical excitation process. \citet{isella2018} reports the CO emission to originate from geometrically thin layers located at a distance $z_{CO} = \pm 30\times(r/100)^{0.5}$ au from the midplane. These authors observe an almost linear decrease of the temperature with the orbital radius in the region comprised between 30 and 500 au, with a drop in temperature within 30 au when moving toward the star that they interpret as resulting from beam dilution \citep{salinas2018}.

Modelling efforts by \citet{isella2018} to reproduce the temperature profile of these geometrically thin CO layers can fit the observations between $\sim$30-50 au and $\sim$280-300 au, but result in predicted temperatures that are too high beyond 280-300 au (or, conversely, the measured temperatures are lower than the expected ones). The orbital region between 30 au 50 au is the one where the geometrically thin CO layers identified by \citet{isella2018} start getting crossed by high-inclination, supersonic planetesimals that can contribute to the heating of the gas (see Figs. \ref{figure-comparison} and \ref{figure-layers}).

Conversely, in our simulations the region extending between 250-300 au is the one where the peak inclination and the density of high-inclination planetesimals experience a significant drop, limiting their possible heating effects on the gas of the outermost part of the circumstellar disk (see Figs. \ref{figure-comparison} and \ref{figure-layers}). While this qualitative spatial match could be purely incidental, it supports the need for further investigations of the effects of supersonic planetesimals on the thermal environment of circumstellar disks.
}

\begin{figure}
\centering
\includegraphics[width=\columnwidth]{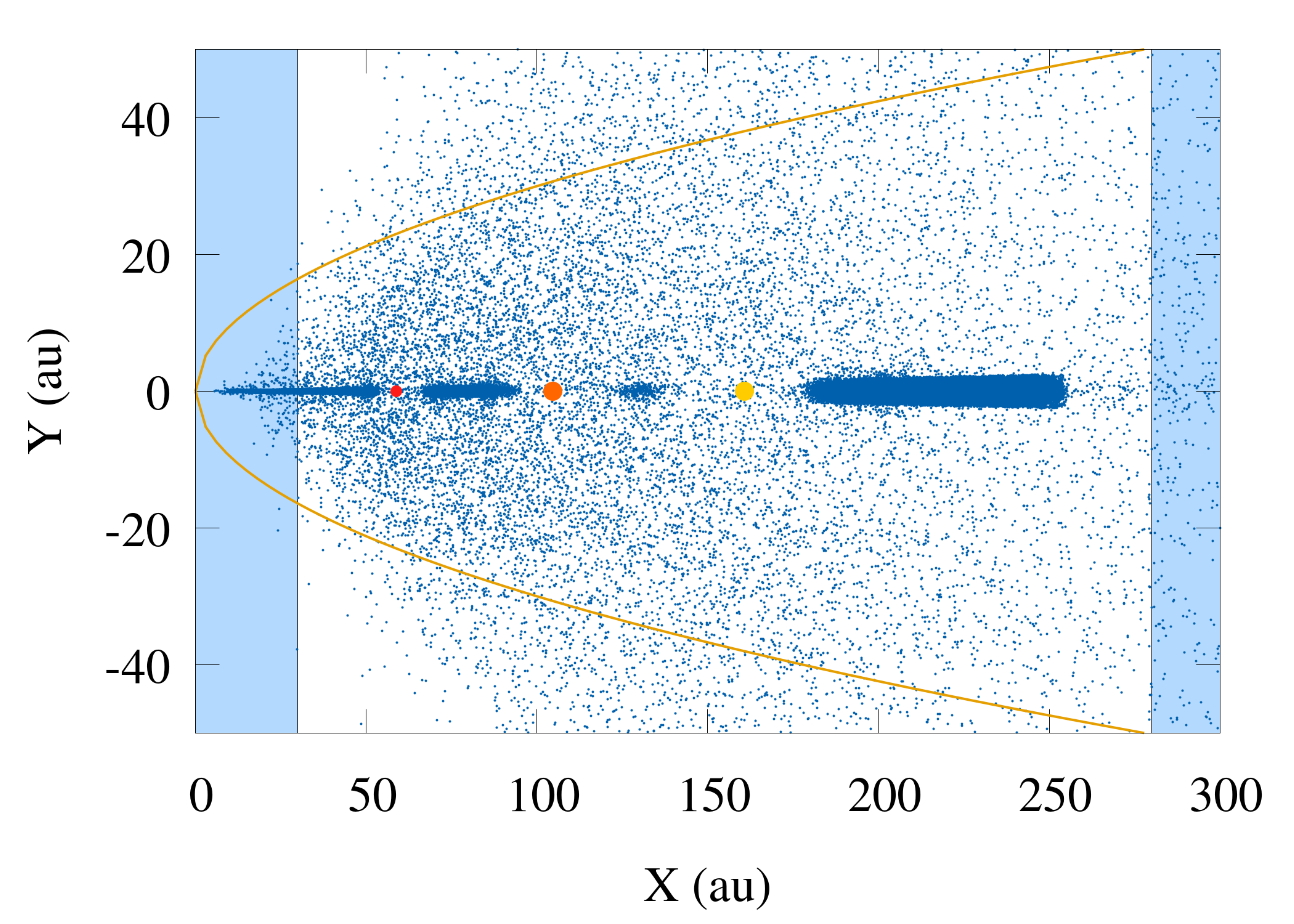}
\caption{{ ``Edge-on'' view of the planetesimal disk at 5 Myr in the ``high mass'' scenario (inner planet with the mass value estimated by \citealt{liu2018}, outer two planets possessing the mass values estimated by \citealt{teague2018}). The two continuous lines indicate the locations of the geometrically thin layers responsible for the CO thermal emission identified by \citet{isella2018}. The two highlighted areas between 0 and 30 au and beyond 280 au are those where the measured temperatures are significantly lower than those predicted by fitting models of the CO-emitting layers and they coincide with the regions where the CO-emitting region is not crossed by supersonic planetesimals (0-30 au) and where the spatial density of supersonic planetesimals drops significantly ($>$280 au).}}\label{figure-layers}
\end{figure}



\section{Discussion and conclusions}

{
The goal of this study was twofold. On one hand, we set out to investigate the general process of dynamical and collisional excitation triggered by the formation of giant planets in circumstellar disks, to assess whether it could be responsible for injecting significant amounts of second--generation dust and affecting the overall dust--to--gas ratio as proposed by \citet{turrini2012}. On the other hand, we wanted to test the effects of this dynamical and collisional excitation process on our specific test bench, HD\,163296's circumstellar disk, to verify whether it could explain its characteristics.

Due to the number of unknown parameters in the properties and evolution of HD\,163296's circumstellar disk and planetesimal population, and in the orbital evolution of the giant planets, we restricted our study to the case of their in situ formation and assumed that migration did not play a major role in their dynamical history. We tested the dependence of the dynamical and collisional excitation process on the planetary masses by considering scenarios encompassing the whole range of proposed values.

To estimate the dust production efficiency of the dynamical excitation process, we developed a simplified collisional model and applied it to the different scenarios we considered. While the description of the collisional environment is based on consolidated statistical methods and up-to-date scaling laws, our model was meant for a first exploration of these processes and includes a number of simplifying assumptions. Future studies with more refined collisional models are therefore warranted.

In this work we considered two possible end-members for the population of planetesimals as described in \citet{krivov2018}: a primordial size-frequency distribution dominated by large planetesimals and a collisionally evolved size-frequency distribution possessing an extended population of small, sub-km planetesimals. The first one represents a realistic starting condition for the planetesimal disk, while the second one a realistic evolved state. 

During the life of HD\,163296's circumstellar disk { the size-frequency distribution of} the planetesimal population should { transform from the primordial one to the collisionally evolved one}. This means that the real collisional dust production is expected to fall somewhere between those computed adopting these two end-member size-frequency distributions.}

The results of our investigation indicate that the formation of HD\,163296's giant planets can indeed cause a late dust--to--gas ratio resurgence in the circumstellar disk by triggering a phase of dynamical excitation of its planetesimal population, halting the steady decay of dust by creating second--generation grains in high-velocity collisions. { It is worth pointing out that our results are not qualitatively affected by the use of HD\,163296's pre-Gaia distance instead of the post-Gaia one. 

While the new, lower distance of the star translates in semimajor axes and stellar mass about $\sim$20$\%$ lower,  the dynamical evolution of the planetesimals in such outer orbital regions remains dominated by the gravitational perturbations of the giant planets. The resulting more compact system, moreover, would be characterized by higher spatial densities of the planetesimals and shorter orbital periods, two effects that would concur in increasing the planetesimal impact probability. As a consequence, our results provide a lower limit to the collisional production of the second-generation dust.}

{ When using the most recent estimates of the planetary masses \citep{liu2018,teague2018}, our collisional model indicates that the dynamical excitation process is always capable of explaining the dust overabundance arising from the results of \citet{isella2016}. If the size-frequency distribution of the planetesimals is similar to the collisionally evolved one we considered, the collisional dust production can actually be responsible for a large fraction, if not the entirety, of the current dust in HD\,163296's circumstellar disk. 

While the detailed reconstruction of the spatial distribution and temporal evolution of the dust--to--gas ratio is beyond the scope of our collisional model, our results indicate that the peak of dust production should occur inside the orbit of the innermost planet due to the combination of the higher spatial density of the planetesimals and of the higher impact velocities (see Figs. \ref{figure-dust-space} and \ref{figure-velocity-comparison}). This is in agreement with the enhanced dust--to--gas ratio found by \citet{isella2016} in the same orbital region with respect to what dynamical models of the disk containing only gas and dust perturbed by the giant planets would predict (see their Fig. 2, right panel). 

Based on the results of \citet{isella2016} and, particularly, on the gas and dust density profiles they reconstructed, about $70$ M$_{\oplus}$ should be injected into this orbital region to explain the observations. According to our collisional model, the dynamical excitation process can produce the required amount of second--generation dust in 1-3 Myr depending on the specific planetary masses and size-frequency distribution of the planetesimals.

{ A further comparison with the masses of the two dust rings located between the three giant planets recently estimated by \citet{dullemond2018} reveals that the collisional dust production process can explain the mass of the inner ring and about 10-20$\%$ the mass of the outer one as the result of 2-3 Myr of collisional evolution of the planetesimals in the case of their collisionally evolved size-frequency distribution. In the case of their primordial size-frequency distribution, instead, over the same timespan the collisional dust production could explain only up to 10$\%$ of the estimated masses of the two rings.

As a consequence, based on our results one could argue that the inner regions of HD\,163296's circumstellar disks (inside the inner giant planet and likely in the ring between the inner and central ones) are dominated by second-generation dust produced by planetesimal collisions, while in the outer regions of the disk (from the ring between the central and outer giant planets outward) the dust population is characterized by a mixture of primordial and second-generation dust, with the formed likely dominating in mass.}

In our investigation, { however}, we did not include the presence of the recently proposed outermost fourth giant planet \citep{pinte2018}, since its orbital and physical characteristics are still loosely constrained. Its proposed mass ($\sim 2$ Jovian masses), moreover, opens up the possibility of its formation having occurred by disk instability instead of core accretion. As such, its role in the evolution of HD\,163296's disk requires a dedicated investigation that will be the subject of future work. 

Based on the dynamical picture arising from our results, however, we can already speculate that the presence of this giant planet would contribute to exciting the outermost regions of the planetesimal disk, raising { their} comparatively lower impact velocities. This would increase the collisional dust production in these region and plausibly provide a better fit to the spatial distribution of the dust-to-gas {{ratio}} as reconstructed by \citet{isella2016} { as well as to the masses of the dust rings estimated by \citet{dullemond2018}, particularly in the case of the primordial size-frequency distribution of the planetesimals}. 

{ Finally, while our simplified collisional model cannot provide information on the vertical spatial distribution of the dust production, the existence of a population of excited planetesimals on high-inclination orbits indicates that, due to the conservation of the angular momentum, part of the dust released by impacts will also be on high-inclination orbits and reside outside of the midplane.

Depending on the balance between the dust production rate and the vertical settling time of these high-inclination dust grains, the dynamical excitation process could replenish and sustain the dust population outside the midplane. This would be in qualitative agreement with one of the explanations proposed for the polarimetric features observed by \citet{guidi2018} in HD\,163296's disk.}

Alongside the \emph{enhanced dust production}, our results also raise the possibility for additional environmental effects of this dynamical and collisional excitation process in the gas of HD\,163296's circumstellar disk. First, most energetic impacts could cause the sublimation of the icy component of the planetesimals and \emph{release transient, non-equilibrium gas species} { like H$_{2}$O, CO$_2$ (as observed during the Deep Impact experiment on comet Tempel 1, \citealt{ahearn2005}), and NH$_{3}$} in the disk, in qualitative agreement with the observations of HD\,163296 and TW Hya by \citet{fedele2012} and \citet{salinas2016} respectively.

{ Furthermore, a collisionally-driven release of CO beyond its snow line would provide an explanation for the DCO+ abundance recently measured by \citet{salinas2018} across the radial extension of  HD\,163296's disk. In particular, the observed radial trend in the abundance of DCO+ (almost constant between 90 and 180 au, with a marked decreasing trend between 180 and 300 au, \citealt{salinas2018}) matches the existence of two different dynamically excited regions, characterized by different impact velocity distributions, over the same orbital region (see Fig. \ref{figure-velocity-comparison}).}

Second, excited planetesimals would move at supersonic speeds with respect to the gas and form bow shocks \citep{weidenschilling1998}. This process could produce observable signatures by { \emph{heating the shocked gas and broadening its emission lines}. The existence of high-inclination, supersonic planetesimals revealed by our simulations suggests the possibility that this process could be acting also outside the midplane and could contribute to the reconstructed CO thermal emission profile of HD\,163296 \citep{isella2018}. The bow shocks created by the supersonic planetesimals} may also contribute to the dust regeneration and the release of non-equilibium gas species by ablating the icy surfaces of the planetesimals, as suggested by the results of \citet{tanaka2013}. Both these effects will be explored in future works.

Finally, due to the first principles approach of this exploratory study, our results highlight how the dynamical excitation process and its associated collisional dust production do not depend on any specific or ad-hoc assumption, but are a natural by-product of the formation of giant planets. As such, these processes should be common to all circumstellar disks in which giant planets form at an early stage of the disk evolution and perturb nearby planetesimals. 

The collisional production of second--generation dust in circumstellar disks hosting giant planets therefore likely represents a common evolutionary phase marking the transition from a circumstellar disk dominated by primordial dust to a debris disk dominated by second--generation dust. Whether the amount of collisionally produced dust is high enough to produce observable signatures, like our results suggest being the case for HD\,163296's disk (and as discussed by \citealt{gratton2019} as a possible explanation for some of the features observed in the disk of HD\,169142), depends on the characteristics of each specific system, first of all the masses of the giant planets and of the planetesimal disk.

As a result, the time dependence of the dust--to--gas ratio on the stellar age may not be a simple linear decay due to dust coagulation into larger bodies and inward drift (see e.g. \citealt{testi2014,pascucci2016}), but it may show sudden bumps related to the formation of giant planets and their interaction with the planetesimal disk. As our results demonstrate, this interaction has the potential of producing effects capable of altering the environment of the circumstellar disks and their observational features.}

\acknowledgments

The authors wish to thank the anonymous referee for the insightful comments that helped improve and expand this work. { They also thank Davide Fedele, Raffaele Gratton, Maria Teresa Capria, Stavro Ivanovski, Eugenio Schisano and Sergio Molinari for the discussions on the dynamical excitation process and its observational implications, and Romolo Politi, Scige John Liu and Mirko Riazzoli for assistance with the computational resources. This work was supported by the project PRIN-INAF 2016 \emph{The Cradle of Life - GENESIS-SKA} (General Conditions in Early Planetary Systems for the rise of life with SKA), by the Italian Ministero dell'Istruzione, Universit\`a e Ricerca (MIUR) through the grant \emph{Progetti Premiali 2012 – iALMA} (CUP C52I13000140001), by the Deutsche Forschungs-gemeinschaft (DFG, German Research Foundation) - Ref no. FOR 2634/1 TE 1024/1-1, and by the DFG cluster of excellence \emph{Origin and Structure of the Universe} (\url{www.universe-cluster.de}).} Additional computational resources were supplied by the INAF-IAPS projects \emph{HPP - High Performance Planetology} and \emph{DataWell}. This research has made use of the Wolfram Alpha ``Computational Intelligence'' service and of the NASA Astrophysics Data System Bibliographic Services.



\end{document}